\documentclass[twocolumn,showpacs,prl]{revtex4}
\usepackage{epsfig,amsfonts,amsbsy}
\newcommand{\ret}{\nonumber\\}

\newcommand{\norm}[1]{\left\Vert#1\right\Vert}

\newcommand{\sbkt}[1]{\langle#1\rangle}

\newcommand{\sumtwo}[2]%
{\mathop{\sum_{#1}}_{#2}}
\newcommand{\sumthree}[3]%
{\mathop{\mathop{\sum_{#1}}_{#2}}_{#3}}
\newcommand{\sumfour}[4]%
{\mathop{\mathop{\mathop{\sum_{#1}}_{#2}}_{#3}}_{#4}} 
\newcommand{\suptwo}[2]%
{\mathop{\sup_{#1}}_{#2}}
\newcommand{\supthree}[3]%
{\mathop{\mathop{\sup_{#1}}_{#2}}_{#3}}
\newcommand{\supfour}[4]%
{\mathop{\mathop{\mathop{\sup_{#1}}_{#2}}_{#3}}_{#4}} 
\newcommand{\inftwo}[2]%
{\mathop{\inf_{#1}}_{#2}}
\newcommand{\infthree}[3]%
{\mathop{\mathop{\inf_{#1}}_{#2}}_{#3}}
\newcommand{\inffour}[4]%
{\mathop{\mathop{\mathop{\inf_{#1}}_{#2}}_{#3}}_{#4}} 
\newcommand{\mintwo}[2]%
{\mathop{\min_{#1}}_{#2}}

\newcommand{\calB}{{\cal B}}

\newcommand{\calN}{{\cal N}}

\newcommand{\up}{\uparrow}
\newcommand{\dn}{\downarrow}

\newcommand{\La}{\Lambda}
\newcommand{\tiLa}{\tilde{\Lambda}}
\newcommand{\HHC}{\mathcal{H}^\mathrm{HC}}
\newcommand{\HLF}{\mathcal{H}^\mathrm{LF}}
\newcommand{\Hup}{\widetilde{\mathcal{H}}^\uparrow}

\newcommand{\Stot}{S_{\rm tot}}
\newcommand{\EGS}{E_\mathrm{GS}}
\newcommand{\Ne}{N_{\rm e}}
\newcommand{\Nmax}{N_\mathrm{max}}
\newcommand{\Nmin}{N_\mathrm{min}}
\newcommand{\Heff}{H_\mathrm{eff}}
\newcommand{\tiA}{\tilde{A}}
\newcommand{\ep}{\epsilon}

\newcommand{\HA}{\tilde{H}_A}
\newcommand{\tH}{\tilde{H}}
\newcommand{\HLa}{\tilde{H}_{\La}}

\newcommand{\vtau}{\boldsymbol{\tau}}
\newcommand{\vsigma}{\boldsymbol{\sigma}}
\newcommand{\eqreff}[1]{(\ref{#1})}
\begin{document}
\title{Metallic ferromagnetism in the Hubbard model: A rigorous example}
\author{Akinori Tanaka${}^{1,\dagger}$ and Hal Tasaki${}^{2,\ast}$}
\affiliation {${}^1$Department of General Education,
Ariake National College of Technology, Omuta, 
Fukuoka 836-8585, Japan
\\
${}^2$Department of Physics, Gakushuin University,
Mejiro, Toshima-ku, Tokyo 171-8588, Japan}

\date{\today}

\vspace{5in}

\begin{abstract}
We present the first rigorous example of the Hubbard model in any dimensions which exhibits metallic ferromagnetism.
The model is a genuine Hubbard model with short-range hopping and
 on-site Coulomb repulsion, 
and has multi single-electron bands.
In the limit where the band gap and the Coulomb repulsion become infinite, we prove that the ground states are completely ferromagnetic and at the same time conducting.
\end{abstract}


\maketitle
It has been believed since Heisenberg \cite{Heisenberg} that  ferromagnetism observed in nature is generated by quantum effects and Coulomb interaction between electrons.
It is a challenging problem to confirm this scenario by showing that  only short-range hopping of electrons and spin-independent Coulomb interaction can lead to ferromagnetism in the concrete setting of the Hubbard model  \cite{review}.

Now many rigorous examples of ferromagnetism (or ferrimagnetism \cite{Lieb}) 
in the Hubbard model are known, and it is clear that certain versions of the model do generate ferromagnetism. 
An important class of examples, now called {\em flat-band ferromagnetism}\/, was discovered by Mielke \cite{Mielke} and then by Tasaki \cite{flat}.
In these models electrons occupy the lowest dispersionless band, and infinitesimally small Coulomb interaction can lead to a complete ferromagnetism.
Related models were found in \cite{more}.
Although the flat-band models are singular in the sense that the
single-electron ground states have huge degeneracy, the mechanism which generates ferromagnetism is believed to be robust and physically realistic.
Indeed the existence of ferromagnetism has been proved rigorously in related nonsingular models  \cite{nonflat,TanakaKagome}.

A common feature of all these rigorous examples of ferromagnetism is that they describe insulators \cite{Nagaoka}.
Metallic ferromagnetism, in which same electrons contribute both to magnetism and conduction, is clearly more interesting and challenging.
As far as we know the only rigorous example of metallic ferromagnetism
in the Hubbard model is that by Tanaka and Idogaki \cite{Tanakametal}, who treated a
quasi one-dimensional model using the Perron-Frobenius argument \cite{review}.
But the physics of one-dimensional electron systems is very special, and it is highly desirable to have examples in higher dimensions.

In this Letter, we present the first rigorous example of metallic
ferromagnetism in a version of the Hubbard model in any dimensions.
The mechanism of ferromagnetism in the present model is basically the
same as that in the previous models, namely, {\em  when one represents
the system using a moderately localized basis, the Coulomb repulsion (in
real space) generates both a repulsive interaction and a ferromagnetic exchange interaction}\/.
Our model is a variant of the models in \cite{flat,nonflat,more} 
and has multi single-electron bands, among which the lowest two mainly contribute to low energy physics (especially in the large band gap limit that we take).
In the ground states the lowest band is half filled 
and exhibits ferromagnetism as in \cite{flat,nonflat,more}.
The electrons in the second lowest band, which is partially filled,
are movable and are coupled  ferromagnetically to the
electrons in the lowest band.
This gives rise to ground states which are ferromagnetic  and at the same time conducting.

Although the basic mechanisms are similar, the mathematical methods developed for the insulating models \cite{flat,nonflat,more} never apply to conducting systems \cite{method}.
We here develop a completely different variational argument.

\begin{figure}
\centerline{\epsfig{file=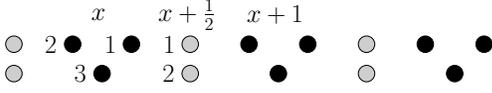,width=6.5cm}}
\caption[dummy]{
The lattice $\tiLa$ for $d=1$.
Integer sites are triplicated, and half-odd-integer sites are duplicated.
}
\label{f:lattice}
\end{figure}

\paragraph*{Definitions and main results:}
Let $\La$ be the $d$-dimensional $L\times\cdots\times L$ hypercubic lattice (where $L$ is even) with unit  lattice spacing and periodic boundary conditions.
Let $\calN=\{(1/2,0,\ldots,0),\ldots,(0,\ldots,0,1/2)\}$ be the set of $d$ vectors of length $1/2$ pointing in the positive direction of each axis.
Then $\calB=\{x+\delta\,|\,x\in\La,\ \delta\in\calN\}$ can be regarded as the set of mid-points of bonds in $\La$.
We construct a Hubbard model  on the lattice $\tiLa=\La\times\{1,2,3\}\cup\calB\times\{1,2\}$ (where the triplicated lattice $\La\times\{1,2,3\}$ consists of pairs $(x,i)$ with $x\in\La$ and $i=1,2,3$, and the duplicated lattice $\calB\times\{1,2\}$ consists of pairs $(w,i)$ with $w\in\calB$ and $i=1,2$).
See Fig.~\ref{f:lattice}.
With each site $z\in\tiLa$ and spin index $\sigma=\up,\dn$, we associate the standard fermion operator $c_{z,\sigma}$.

\begin{figure}
\centerline{\epsfig{file=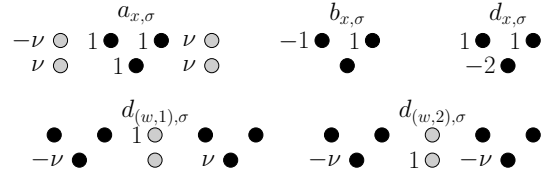,width=7cm}}
\caption[dummy]{
Components of the states corresponding to the special fermion operators  indexed by $x\in\La$ and $w\in\calB$.
We omitted the normalization factors for the $a$ and $b$ operators.
Each state is localized within the unit lattice spacing.
}
\label{f:states}
\end{figure}

It is convenient to define some fermion operators by combining the basic operator $c_{z,\sigma}$.
For each $x\in\La$, $\delta\in\calN$, $i=1,2$ and $\sigma=\up,\dn$, we define
\begin{eqnarray}
a_{x,\sigma}&=&\frac{1}{\sqrt{3+4d\nu^2}}\Bigl[
\sum_{i=1}^3c_{(x,i),\sigma}
\ret
&+&\nu\hspace{-5pt}\sum_{\delta\in\calN,\ i=1,2}
\hspace{-5pt}
\{c_{(x+\delta,i),\sigma}+(-1)^ic_{(x-\delta,i),\sigma}\}
\Bigr],
\label{e:axs}
\\
b_{x,\sigma}&=&\frac{1}{\sqrt{2}}\{c_{(x,1),\sigma}-c_{(x,2),\sigma}\},
\label{e:bxs}
\\
d_{x,\sigma}&=&c_{(x,1),\sigma}+c_{(x,2),\sigma}-2c_{(x,3),\sigma},
\label{e:dxs}
\\
d_{(x+\delta,i),\sigma}&=&c_{(x+\delta,i),\sigma}-
\nu\{c_{(x,3),\sigma}+(-1)^ic_{(x+2\delta,3),\sigma}\},
\label{e:dus}
\end{eqnarray}
where $\nu>0$, is a model parameter, whose value does 
not play essential roles in the present work.
See Fig.~\ref{f:states}.
These operators are designed in such a way that any electronic state on $\tiLa$ can be written by a combination of $a$, $b$, and $d$ operators.
Moreover one has $\{a^\dagger,b\}=\{a^\dagger,d\}=\{b^\dagger,d\}=0$ for any combinations of indices, and $\{a^\dagger_{x,\sigma},a_{y,\tau}\}=\{b^\dagger_{x,\sigma},b_{y,\tau}\}=\delta_{x,y}\delta_{\sigma,\tau}$ for any $x,y\in\La$ and $\sigma,\tau=\up,\dn$.
Note that, unlike in our previous models \cite{flat,nonflat}, the $a$-operators satisfy the standard canonical anticommutation relations.

We define Hamiltonian $H$ by
\begin{eqnarray}
H&=&
\sumtwo{x,y\in\La;\,|x-y|=1}{\sigma=\up,\dn}(-s\,a^\dagger_{x,\sigma}a_{y,\sigma}
-t\,b^\dagger_{x,\sigma}b_{y,\sigma})
\ret
&&
+u\,\Bigl\{\sumtwo{x\in\La}{\sigma=\up,\dn}
d^\dagger_{x,\sigma}d_{x,\sigma}
+\sumtwo{w\in\calB,\,i=1,2}{\sigma=\up,\dn}
d^\dagger_{(w,i),\sigma}d_{(w,i),\sigma}\Bigr\}
\ret
&&
+v\sumtwo{x\in\La}{\sigma=\up,\dn}b^\dagger_{x,\sigma}b_{x,\sigma}
+U\sum_{z\in\tiLa}n_{z,\up}n_{z,\dn},
\label{e:H}
\end{eqnarray}
with $n_{z,\sigma}=c^\dagger_{z,\sigma}c_{z,\sigma}$.
Note that \eqreff{e:H} defines a genuine Hubbard model with short ranged (but admittedly complicated) hopping amplitudes.
The model has several bands; the $a$-band with the dispersion relation $\epsilon_a({\bf k})=-2s\sum_{i=1}^d\cos(k_i)$, the $b$-band with  $\epsilon_b({\bf k})=v-2t\sum_{i=1}^d\cos(k_i)$, and the $d$-bands with higher energies.
We fix the total electron number to $\Ne$.

{\em Theorem---}\/
Let $d=1,2,3,\ldots$ be arbitrary and suppose that $|\La|\le\Ne\le2|\La|$ \cite{set} and $v>2d\,(|s|+2|t|)$.
In the limit $u,U\to\infty$ \cite{Uu}, the ground states of \eqreff{e:H} exhibit saturated ferromagnetism in the sense that they have  the maximum possible total spin $\Stot=\Ne/2$.

One may replace the lower bound for $v$ by better values which depend on the electron number.
For example it is enough to have $v>2d\,(|s|+|t|)$ when $3|\La|/2<\Ne\le2|\La|$.

The electron number $\Ne=|\La|$ corresponds to the half-filling of the lowest $a$-band, and  $\Ne=2|\La|$  to the half-filling of the $a$ and $b$-bands.
Therefore when $t\ne0$ and the electron number satisfies $|\La|<\Ne<2|\La|$, the ferromagnetic ground states, which are indeed Slater determinant states,  are conducting states with $\Ne-|\La|$ conducting electrons (or $2|\La|-\Ne$ holes) in the $b$-band.

\paragraph*{Finite energy states:}
We shall describe a complete proof of the theorem.
We say that $\Phi$ is a finite energy state if $\sbkt{\Phi,H\Phi}<\infty$ in the limit $u,U\to\infty$.
A finite energy state cannot contain any of the $d$-states since a $d$-electron costs an energy proportional to $u$, which becomes infinite.
Furthermore since $U\to\infty$, a finite energy state $\Phi$ must satisfy for any $z\in\tiLa$ the condition $n_{z,\up}n_{z,\dn}\Phi=0$ and hence
\begin{equation}
c_{z,\dn}c_{z,\up}\Phi=0.
\label{e:ccP}
\end{equation}

Let $\Phi_0$ be the state with no electrons.
A computation shows that $c_{(x,3),\dn}c_{(x,3),\up}(\cdots)a^\dagger_{x,\up}a^\dagger_{x,\dn}\Phi_0=(3+4d\nu^2)^{-1}(\cdots)\Phi_0$, where  $(\cdots)$ is an arbitrary product of $a^\dagger$ and $b^\dagger$ except for $a^\dagger_{x,\up}$, $a^\dagger_{x,\dn}$.
This means that any state $\Phi$ which contains a term with $a^\dagger_{x,\up}a^\dagger_{x,\dn}$ cannot satisfy $c_{(x,3),\dn}c_{(x,3),\up}\Phi=0$.
Thus a finite energy state  $\Phi$ has no terms with $a^\dagger_{x,\up}a^\dagger_{x,\dn}$.
Likewise \cite{bb} we can show that $\Phi$ has no terms with $b^\dagger_{x,\up}b^\dagger_{x,\dn}$.
Therefore any finite energy state is in the subspace $\HHC$ with the ``hard core condition'', which  is spanned by the basis states
\begin{equation}
\Psi(A,\vsigma;B,\vtau)=\Bigl(\prod_{x\in A}a^\dagger_{x,\sigma(x)}\Bigr)
\Bigl(\prod_{x\in B}b^\dagger_{x,\tau(x)}\Bigr)
\Phi_0,
\label{e:PsAB}
\end{equation}
where $A$, $B$ are arbitrary subsets of $\La$ such that $|A|+|B|=\Ne$, and $\vsigma=(\sigma(x))_{x\in A}$,  $\vtau=(\tau(x))_{x\in B}$  are arbitrary spin configurations with $\sigma(x),\tau(x)\in\{\up,\dn\}$ \cite{order}.

For a state $\Phi$ to satisfy \eqreff{e:ccP}, it is not enough that $\Phi\in\HHC$.
By imposing \eqreff{e:ccP} for other sites, we find that a finite energy state $\Phi$ must satisfy the following {\em local ferromagnetic conditions}\/.
When we expand $\Phi$ as
\begin{equation}
\Phi=\sum_{A,\vsigma,B,\vtau}\psi(A,\vsigma;B,\vtau)\,\Psi(A,\vsigma;B,\vtau),
\label{e:Phiexp}
\end{equation}
the coefficients $\psi(A,\vsigma;B,\vtau)$ must satisfy $\psi(A,\vsigma;B,\vtau)=\psi(A,\vsigma_{x\leftrightarrow y};B,\vtau)$ for any $x,y\in A$ such that $|x-y|=1$, and $\psi(A,\vsigma;B,\vtau)=\psi(A,\vsigma_x;B,\vtau_x)$ for any $x\in A\cap B$.
Here $\vsigma_{x\leftrightarrow y}$ is the configuration obtained from $\vsigma$ by switching $\sigma(x)$ and $\sigma(y)$ in $\vsigma$.
Similarly $\vsigma_x,\vtau_x$ are obtained by switching $\sigma(x)$ and $\tau(x)$ in $\vsigma,\vtau$~\cite{localferro}.
These conditions are equivalent to {\em infinitely large ferromagnetic couplings between neighboring $a$-electrons, and between the $a$-electron and the $b$-electron sharing a same site $x$.}

By $\HLF$ we denote the subspace of $\HHC$ consisting of states which
satisfy the local ferromagnetic conditions.
Note that for any $\Phi\in\HLF$ the expectation value of $H$ satisfies
$\sbkt{\Phi,H\Phi}=\sbkt{\Phi,\Heff\Phi}$ with
\begin{equation}
\Heff=\sum_{x,y,\sigma}(-s\,a^\dagger_{x,\sigma}a_{y,\sigma}-t\,b^\dagger_{x,\sigma}b_{y,\sigma})+v\sum_{x,\sigma}b^\dagger_{x,\sigma}b_{x,\sigma}.
\label{e:Heff}
\end{equation}

\paragraph*{Variational estimates:}
So far all of the arguments are straightforward variations of those developed for the simple flat-band models \cite{flat}.
Let us now turn to variational estimates, 
which are essential to our treatment of  conducting states.

Note that the  above stated local ferromagnetic conditions relate the coefficients $\psi(A,\vsigma;B,\vtau)$ with common $A$ and $B$.
We can thus decompose $\HLF$ into a direct sum as $\HLF=\bigoplus_{N_a=\Ne-|\La|}^{|\La|}\HLF_{N_a}$.
Here $\HLF_{N_a}$ is the intersection of $\HLF$ and the space spanned by the basis states $\Psi(A,\vsigma;B,\vtau)$ with any $A$ such that $|A|=N_a$, and arbitrary $\vsigma$, $B$, and $\vtau$.
Since the effective Hamiltonian \eqreff{e:Heff} leaves the number of $a$-electrons invariant, we can determine the ground state energy $\EGS$ variationally as
\begin{equation}
E(N_a)=\inftwo{\Phi\in\HLF_{N_a}}{\norm{\Phi}=1}\sbkt{\Phi,\Heff\Phi},\quad
\EGS=
\hspace{-10pt}\mintwo{N_a}{\Ne-|\La|\le N_a\le|\La|}
\hspace{-10pt}
E(N_a).
\label{e:EGS}
\end{equation}

When $N_a=|\La|$, $a$-electrons fill the entire $\La$, and are coupled ferromagnetically.
Since all $b$-electrons are coupled ferromagnetically to the $a$-electrons, we see that any state in $\HLF_{|\La|}$ has the maximum possible total spin $\Stot=\Ne/2$.
It is also easy to see that $E(|\La|)$ gives the lowest energy among the ferromagnetic states.
In what follows, we shall prove that $E(N_a)>E(|\La|)$ for any $N_a<|\La|$.
This shows that the ground states have the maximum total spin, and  proves our theorem.

Let $N_a<|\La|$.
We first note that on the space $\HLF_{N_a}$,
\begin{equation}
\Heff\ge \tH+(\Ne-N_a)\,v-2d|s|\,(|\La|-N_a),
\label{e:HLB1}
\end{equation}
where
\begin{equation}
\tH=-t\sumtwo{x,y\in \La;\,|x-y|=1}{\sigma=\up,\dn}
b^\dagger_{x,\sigma}b_{y,\sigma}.
\label{e:tH}
\end{equation}
To get the lower bound \eqreff{e:HLB1}, we noted that each hole (i.e., 
a site in $\La$ not occupied by an $a$-electron) has a kinetic energy
not less than $-2d|s|$ \cite{Shen}.

Since $\tH$ does not act on  $a$-electrons, we have
\begin{equation}
\inftwo{\Phi\in\HLF_{N_a}}{\norm{\Phi}=1}\sbkt{\Phi,\tH\Phi}
=\mintwo{A\subset\La}{|A|=N_a}
\inftwo{\Phi\in\HLF_A}{\norm{\Phi}=1}\sbkt{\Phi,\tH\Phi},
\label{e:inftH}
\end{equation}
where $\HLF_A$ is the intersection of $\HLF$ and the space spanned by the basis states $\Psi(A,\vsigma;B,\vtau)$ with the specified $A$ and arbitrary $\vsigma$, $B$, and $\vtau$.

Note that, on $\HLF_A$ (even on $\HHC$), we can bound $\tH$ as
\begin{equation}
\tH\ge\HA-2d|t|\,(|\La|-|A|),
\label{e:HLB2}
\end{equation}
for any $A\subset\La$, where 
\begin{equation}
\HA=-t\sumtwo{x,y\in A;\,|x-y|=1}{\sigma=\up,\dn}
b^\dagger_{x,\sigma}b_{y,\sigma}
\label{e:HA}
\end{equation}
is the hopping Hamiltonian restricted on $A$.
To get the lower bound \eqreff{e:HLB2}, we applied the bound $|\sum_\sigma(b^\dagger_{x,\sigma}b_{y,\sigma}+\mathrm{h.c.})|\le1$ (which is valid on $\HHC$) to all the hopping terms including any site in $\La\backslash A$.
From \eqreff{e:EGS}, \eqreff{e:HLB1}, \eqreff{e:inftH} and \eqreff{e:HLB2}, we have
\begin{eqnarray}
E(N_a)&\ge&
(\Ne-|\La|)v+(|\La|-N_a)\{v-2d\,(|s|+|t|)\}
\ret&&
+
\mintwo{A\subset\La}{|A|=N_a}\inftwo{\Phi\in\HLF_A}{\norm{\Phi}=1}
\sbkt{\Phi,\HA\Phi}.
\label{e:EALB}
\end{eqnarray}

We shall examine the infimum in \eqreff{e:EALB}.
Let us decompose $A$ into connected components as $A=\bigcup_{i=1}^n\tiA_i$.
Within each $\tiA_i$, all the $a$-electrons and $b$-electrons are coupled  to have the maximum possible total spin because of the local ferromagnetic conditions.
Note that $\HA$ allows $b$-electrons to hop around only within each
connected component $\tiA_i$, and leaves those $b$-electrons on
$\La\backslash A$ unaffected.
This means that the above ferromagnetic coupling is not disturbed by the application of $\HA$.
Therefore the infimum of the expectation value of $\HA$ taken over all
states in $\HLF_A$  can be estimated simply 
in the subspace spanned by up-spin electrons \cite{replace}.
At this stage we can forget about the $a$-electrons, which have no kinetic energies in $\HA$, and consider only the (now fully polarized) $b$-electrons.
This leads us to
 \begin{equation}
\inftwo{\Phi\in\HLF_A}{\norm{\Phi}=1}\sbkt{\Phi,\HA\Phi}
=\inftwo{\Phi\in\Hup_{\La,\Ne-|A|}}{\norm{\Phi}=1}\sbkt{\Phi,\HA\Phi},
\label{e:HupA}
\end{equation}
where $\Hup_{\La,N}$ is the space spanned by states of the form $(\prod_{x\in B}b^\dagger_{x,\up})\Phi_0$ with an arbitrary $B\subset\La$ such that $|B|=N$.
An inspection shows that, in the space $\Hup_{\La,\Ne-|A|}$, the number of movable electrons (which are on $A$, and hence acted on by $\HA$) varies from $\Nmin=\Ne-|\La|$ to $\Nmax=\min\{|A|,\Ne-|A|\}$.

Since unmovable electrons (which are on $\La\backslash A$) are not affected by $\HA$ at all, we see that $\sbkt{(\prod_{x\in B}b^\dagger_{x,\up})\Phi_0,\HA\,(\prod_{x\in B'}b^\dagger_{x,\up})\Phi_0}=0$ whenever $|A\cap B|\ne |A\cap B'|$.
Therefore we can evaluate the infimum in the right-hand side of \eqreff{e:HupA} in each subspace with a fixed number of movable electrons.
Since unmovable electrons have no contributions to  expectation values of $\HA$, we have
\begin{equation}
\inftwo{\Phi\in\Hup_{\La,\Ne-|A|}}{\norm{\Phi}=1}\sbkt{\Phi,\HA\Phi}
=
\mintwo{N}{\Nmin\le N\le\Nmax}
\inftwo{\Phi\in\Hup_{A,N}}{\norm{\Phi}=1}\sbkt{\Phi,\HA\Phi},
\label{e:mininf}
\end{equation}
where $\Hup_{A,N}$ is the space spanned by states of the form $(\prod_{x\in B}b^\dagger_{x,\up})\Phi_0$ with an arbitrary $B\subset A$ such that $|B|=N$.

Let  $\HLa$ be the hopping Hamiltonian \eqreff{e:HA} with $A=\La$.
Since $\HLa$ and $\HA$ are equivalent when restricted to the subspace $\Hup_{A,N}$, we see that
\begin{equation}
\inftwo{\Phi\in\Hup_{A,N}}{\norm{\Phi}=1}\sbkt{\Phi,\HA\Phi}
=
\inftwo{\Phi\in\Hup_{A,N}}{\norm{\Phi}=1}\sbkt{\Phi,\HLa\Phi}
\ge
\inftwo{\Phi\in\Hup_{\La,N}}{\norm{\Phi}=1}\sbkt{\Phi,\HLa\Phi},
\label{e:infthree}
\end{equation}
where the inequality follows from $\Hup_{A,N}\subset\Hup_{\La,N}$.
We then find, from \eqreff{e:HupA}, \eqreff{e:mininf}, and \eqreff{e:infthree}, that
\begin{equation}
\inftwo{\Phi\in\HLF_A}{\norm{\Phi}=1}\sbkt{\Phi,\HA\Phi}
\ge
\mintwo{N}{\Nmin\le N\le\Nmax}
\inftwo{\Phi\in\Hup_{\La,N}}{\norm{\Phi}=1}\sbkt{\Phi,\HLa\Phi}.
\label{e:mininf2}
\end{equation}

Let $\ep_1\le \ep_2\le\cdots\le\ep_{|\La|}$ be the eigenvalues of the hopping Hamiltonian $\HLa$ (which is \eqreff{e:HA} with $A=\La$) in ascending order.
Since the energy spectrum has a plus-minus symmetry, we see that $\ep_\ell\le0$ if $\ell\le|\La|/2$ and  $\ep_\ell\ge0$ if $\ell>|\La|/2$.
The infimum in the right-hand side of \eqreff{e:mininf2} is nothing but the ground state energy of a spinless free fermion, and is equal to $\sum_{\ell=1}^N\ep_\ell$.
By minimizing this over $N$, we see that \cite{min}
\begin{equation}
(\text{right-hand side of \eqreff{e:mininf2}}) \ge(|\La|-|A|)\,\bar{\ep}
+\sum_{\ell=1}^{\Nmin}\ep_\ell,
\label{e:below2}
\end{equation}
where $\bar{\ep}$ is 0 if $\Nmin>|\La|/2$, and is $\ep_{\Nmin}$ if $\Nmin\le|\La|/2$.

As for the lowest energy $E(\La)$ of the ferromagnetic states, one has
\begin{equation}
E(|\La|)=(\Ne-|\La|)\,v+\sum_{\ell=1}^{\Ne-|\La|}\ep_\ell,
\label{e:ELa}
\end{equation}
since one simply fills all the $a$-states and the lowest $\Ne-|\La|$ states of the $b$-band with up-spin electrons to get the lowest energy.
Note that the total energy of the fully filled $a$-band is vanishing since there are no diagonal terms in the hopping Hamiltonian of the $a$-electrons.

Combining \eqreff{e:EALB}, \eqreff{e:mininf2},  \eqreff{e:below2}, and \eqreff{e:ELa}, we finally get
\begin{equation}
E(N_a)\ge E(|\La|)+(|\La|-|A|)\,\{v-2d\,(|s|+|t|)+\bar{\ep}\},
\label{e:EAEL}
\end{equation}
which implies the desired bound $E(N_a)>E(|\La|)$ if $N_a<|\La|$ and $v>2d\,(|s|+|t|)-\bar{\ep}$.
Since $\bar{\ep}\ge-2d|t|$, we get the condition for $v$ in the theorem.
The improved condition is obtained by recalling that $\bar{\ep}=0$ if $\Nmin>|\La|/2$.

We wish to thank Teppei Sekizawa for useful discussions which, after a few years, led us to an essential observation.
A.T. is supported by Grant-in-Aid for Young Scientists (B), (18740243), from
MEXT, Japan.


\begin{thebibliography}{10}
\bibitem{method}
The basic strategy was to decompose the total Hamilton $H$ into a sum of (mutually non-commuting) local Hamiltonians as $H=\sum h_x$, and to look for ground states which minimize every local Hamiltonian $h_x$.
The resulting ground states inevitably have small charge fluctuation, and are hence insulating.
\bibitem{set}
By $|S|$ we denote the number of elements in a set $S$.
\bibitem{Uu}
Although we believe that the theorem is valid for sufficiently large but finite $U$ and $u$ for any $L$, the presently available techniques are insufficient for proving this.
\bibitem{bb}
Let $(\cdots)$ be an arbitrary product of $a^\dagger$ and $b^\dagger$ for sites other than $x$.
Then we have $c_{(x,1),\dn}c_{(x,1),\up}(\cdots)b^\dagger_{x,\up}b^\dagger_{x,\dn}\Phi_0=(1/2)(\cdots)\Phi_0$, and $\sum_{i=1,2}c_{(x,i),\dn}c_{(x,i),\up}(\cdots)a^\dagger_{x,\sigma}b^\dagger_{x,\up}b^\dagger_{x,\dn}\Phi_0=(\cdots)a^\dagger_{x,\sigma}\Phi_0$.
These show the claim.
\bibitem{order}
We order the sites in $\La$ in an arbitrary but a fixed manner.
Products 
are  arranged according to this order.
\bibitem{localferro}
Let $x,y\in\La$ with $|x-y|=1$ and set $w=(x+y)/2$.
If $(\cdots)$ is an arbitrary product of $a^\dagger$ and $b^\dagger$ except for $a^\dagger_{x,\up}$, $a^\dagger_{x,\dn}$, $a^\dagger_{y,\up}$, $a^\dagger_{y,\dn}$, we have
$c_{(w,2),\dn}c_{(w,2),\up}(\cdots)a^\dagger_{x,\sigma}a^\dagger_{y,\tau}\Phi_0=\nu^2(3+4d\nu^2)^{-1}(\cdots)\{\delta_{\up,\sigma}\delta_{\dn,\tau}-\delta_{\dn,\sigma}\delta_{\up,\tau}\}\Phi_0$.
This, when applied to the expansion \eqreff{e:Phiexp}, yields the first condition on $\psi$.
Likewise, if $(\cdots)$ is an arbitrary product of $a^\dagger$ and $b^\dagger$  for sites other than $x$, we have
$c_{(x,1),\dn}c_{(x,1),\up}(\cdots)a^\dagger_{x,\sigma}b^\dagger_{x,\tau}\Phi_0=\{2(3+4d\nu^2)\}^{-1/2}(\cdots)\{\delta_{\up,\sigma}\delta_{\dn,\tau}-\delta_{\dn,\sigma}\delta_{\up,\tau}\}\Phi_0$,
which leads to the second condition on $\psi$.
\bibitem{replace}
Let 
$\Phi_\up$ be any normalized state in $\HLF_A$ 
which consists only of up-spin electrons.
Let $S^-_x=a^\dagger_{x,\dn}a_{x,\up}+b^\dagger_{x,\dn}b_{x,\up}$ and $S^-_i=\sum_{x\in\tilde{A}_i}S^-_x$ be the spin lowering operators.
Define 
  $S({\bf n})
     =\prod_{x\in\La\backslash A}(S^-_x)^{n_x}\prod_{i=1}^n(S^-_i)^{n_i}$ 
with $n_x=0,1$ and $n_i=0,1,2\ldots$, 
where ${\bf n}$ is a short hand for a collection of these $n$'s,
and define a normalized state 
$\Phi_{\bf n}
=(\text{normlization})\,S({\bf n})\Phi_{\up}$.
Then the space $\HLF_A$ is spanned by the states of the form $\Phi_{\bf n}$.
Since $\HA$ and $S({\bf n})$ commute, 
$\sbkt{\Phi_{{\bf n}^\prime},\HA\Phi_{\bf n}}$
equals
$\sbkt{\Phi_{\up},\HA\Phi_{\up}}$ if ${\bf n}^\prime={\bf n}$
and is zero otherwise. Thus the claim follows.
\bibitem{min} 
Suppose that 
$\Nmin>|\La|/2$.
Then since $\ep_\ell\ge0$ for $\ell\ge\Nmin$, we see that the desired minimum is $\sum_{\ell=1}^{\Nmin}\ep_\ell$.
When 
$\Nmin\le|\La|/2$,
we can bound the  minium from below by
$\sum_{\ell=1}^{\Nmin}\ep_\ell+(\Nmax-\Nmin)\,\ep_{\Nmin}
\ge \sum_{\ell=1}^{\Nmin}\ep_\ell+(|\La|-|A|)\,\ep_{\Nmin}$,
where we used $\Nmin=\Ne-|\La|$ and $\Nmax\le\Ne-|A|$.
\end{thebibliography}
\end{document}